\documentstyle[12pt]{article}
\def\t{\theta}
\def\o{\over}
\def\l{\lambda}
\def\g{\gamma}
\def\s{\sigma}
\def\a{\alpha}
\def\b{\beta}
\def\om{\omega}
\def\Th{\Theta}
\def\T{{\cal T}}
\def\V{{\cal V}}
\def\H{{\cal H}}
\def\G{{\cal G}}
\def\CC{{\rm | \! \!  C}}
\def\Im{{\rm Im\,}}
\def\bra#1{\left\langle #1\right|}               
\def\ket#1{\left| #1\right\rangle}               

\begin{document}
\begin{titlepage}
\title{\bf THE FIELD THEORY LIMIT \\
 OF INTEGRABLE LATTICE MODELS}
\author{ H.J. de Vega\dag \\
{\it  \dag  Laboratoire de Physique Th\'eorique et Hautes Energies,}\\
{\it Universit\'e Pierre et Marie Curie (Paris VI)}\\
 {\it et Universit\'e Denis Diderot (Paris VII), }\\
 {\it  Tour 16, 1er. \'etage, 4, Place Jussieu} \\
 {\it 75252 Paris, cedex 05, France.} \\
 {\it Laboratoire Associ\'{e} au CNRS URA280.}}
\date{June 1994}
\maketitle
\begin{abstract}
The light-cone approach is reviewed. This method allows to find the
underlying quantum field theory for any integrable lattice model in its
gapless regime. The relativistic spectrum and S-matrix follows
straightforwardly in this way through the  Bethe Ansatz.
We show here  how to derive the infinite number of local commuting
and non-local and
non-commuting conserved charges in integrable QFT, taking the massive
Thirring model (sine-Gordon) as an example.
They are generated by quantum monodromy operators and provide a
representation of $q-$deformed affine Lie algebras $U_q({\hat\G})$.
\end{abstract}

\vskip-16.0cm
\rightline{{\bf LPTHE--PAR 94/26}}
\rightline{{\bf June 1994}}
\vskip2cm
\end{titlepage}


\begin{section}{\bf Yang-Baxter equations and the Light-cone Approach }

How to take the continuum limit of integrable lattice models has
always been a major problem.

In this short review we shall try to convince the reader that the
light-cone approach \cite{npb,jpa}
is the best way to perform such continuum limit.

We consider a $ N $ x $ M $ two dimensional square lattice whose links are
labeled by an index $ a = 1, \ldots, n$. The statistical weight of
each vertex where the four links meet is defined by the $R-$matrix $
R^{ab}_{cd}(\t) $ where $ 1 \leq a,b,c,d \leq n$ (see fig.1).
Here $ \theta $ is a complex variable called spectral variable. In the
present context  $\t$ can be considered as a sort of coupling constant. It
must be noticed that  universal magnitudes are
 $ \theta $-independent in integrable models   \cite{ij}.

It is convenient to introduce an operator $ T_{ab}(\t ,{\tilde \omega}) $
associated to horizontal lines (see fig. 2)

\begin{equation}
T_{ab}(\t,{\tilde \omega})= \sum_{a_1,...,a_{N-1}} t_{a_1b}(\t +
\omega_1 ) \otimes  t_{a_2a_1}(\t + \omega_2 )\otimes .....
\otimes  t_{aa_{N-1}}(\t + \omega_N )
\label{monin}
\end{equation}
For fixed $a,b,~ T_{ab}(\theta, {\tilde \omega}) $ acts on the vertical space
$\V  = \bigotimes_{1 \leq i \leq N}
V_i ~~, V_i \equiv  \CC^n$
 , and the local vertex operators  are defined as  $\left[
t_{ab}(\t)\right]_{cd} \equiv R^{bd}_{ca}(\t) $.
 In eq.(\ref{monin}) we introduced  arbitrary inhomogeneity parameters
 $ \omega_1 , \omega_2 ,...., \omega_N $ associated to each site on
a horizontal line.

When periodic boundary conditions (PBC) are considered, it is useful
to define the row-to-row transfer matrix as
\begin{equation}
t(\t,{\tilde \omega})=\sum_a \; T_{aa}(\t,{\tilde \omega})
\label{trans}
\end{equation}
For other types of boundary conditions, see refs.\cite{ij,skl,ale}.

The first  relevant physical problem is to compute the partition
function ${\cal Z}$. It is defined as the sum over all possible
configurations of the statistical weights for the whole lattice.
For a  $ N $ x $ M $ lattice with PBC in both directions,  ${\cal Z}$ can
be written as
\begin{equation}
 {\cal Z} = Tr\left[ t(\t,{\tilde \omega})^M \right]
\label{elib}
\end{equation}

where $ Tr $ stands for the trace on the vertical space ${\cal  V}$.
The free energy is then given by
 \begin{equation}
f(\t,{\tilde \omega}) = - \lim_{(N,M)\to\infty}\;{1\over{NM}}\log{\cal Z}
\label{enli}
\end{equation}
All considerations up to now are valid whether the model is integrable
or not. We shall call integrable those models where the $R$-matrix
$R(\t)$ obeys the Yang-Baxter equations (YBE):
\begin{eqnarray}
\sum_{1\leq k,l,m \leq n} & R^{kl}_{ba}(\t-\t') R^{dm}_{ck}(\t)
R^{ef}_{ml}(\t') =\nonumber\\
\sum_{1\leq k,l,m \leq n} & R^{lk}_{cb}(\t') R^{mf}_{ka}(\t)
R^{de}_{lm}(\t-\t') \label{ybe}
\end{eqnarray}
or in tensor product notation
\begin{eqnarray}
& \left[ 1 \otimes R(\t-\t') \right] \left[ R(\t) \otimes 1\right]
\left[ 1  \otimes R(\t') \right] = \nonumber \\
& \left[  R(\t')\otimes 1\right]\left[ 1  \otimes R(\t) \right]
 \left[ R(\t-\t') \otimes  1 \right] \label{ybee}
\end{eqnarray}
It must be stressed that the YBE are a heavily overdetermined set of
functional algebraic equations. They contain {\it a priori} $n^4$
unknowns [the elements of $R(\t)$] and $n^6$ equations. Despite this
fact a large set of solutions is known. All of them possess
symmetries that reduce the number of independent equations and make
possible the existence of solutions. The symmetries may be discrete as
cyclic $Z_n$ symmetries, continuous abelian symmetries as $U(1)^n$ and
non-abelian as $GL(n)$ (see \cite{ij}).

The YBE (\ref{ybe}-\ref{ybee}) admit the natural graphical representation
given in fig. 3. Graphically, the YBE express the freedom to push
lines through intersections of pair of lines. This possibility of
rigid line shifting can be interpreted as a zero curvature condition on
the lattice.

The YBE   enjoy a powerful coproduct property. Namely,
 eqs.(\ref{ybe}-\ref{ybee}) implies that the operators  $T_{ab}(\t,{\tilde
\omega})$  fulfill the YB algebra
\begin{equation}
 R(\l-\mu)\left[T(\l,{\tilde \omega} )\otimes
     T(\mu,{\tilde \omega} )\right] =\left[T(\mu,{\tilde \omega} )\otimes
     T(\l,{\tilde \omega} )\right]R(\l-\mu)
 \label{nuda}
\end{equation}
For one-site $(N=1), T_{ab}(\t,{\tilde\omega})$ reduces to $R(\t)$ and
eq.(\ref{nuda}) becomes eq.(\ref{ybe}). For $N$-sites,
eq.(\ref{nuda})  can be easily proved by  repeatedly pushing lines
through vertices (see \cite{ij}).
That is, eq.(\ref{nuda}) is the expression of the YBE for $N$-sites.
The coproduct rule is here defined by  eq.(\ref{monin}).
Eq.(\ref{nuda}) implies the commutativity of transfer matrices
\begin{equation}
\left[t(\t,{\tilde \omega}) ,t(\t',{\tilde \omega})\right] = 0
\label{comut}
\end{equation}
That is, the transfer matrices {\it for fixed}
  $ \omega_1 , \omega_2 ,...., \omega_N $ form a commuting family.
Hence, one can expect to diagonalize it with $\t$-independent
eigenvectors:
\begin{equation}
t(\t,{\tilde \omega}) \Psi({\tilde \omega}) = \Lambda(\t,{\tilde
\omega})~ \Psi({\tilde \omega})
\label{max}
\end{equation}
The Bethe Ansatz (BA) actually {\bf does} this job \cite{ij}.
Then, the free energy in the thermodynamic limit turns to be  given
by the largest eigenvalue $\Lambda(\t,{\tilde\omega})_{max}$ of
$t(\t,{\tilde \omega})$. We find from eqs.(\ref{enli}) and (\ref{max})
\begin{equation}
f(\t,{\tilde \omega}) = -
\lim_{N\to\infty}\;{1\over{N}}\log\Lambda(\t,{\tilde\omega})_{max}
\label{enlif}
\end{equation}

When $\t = \t'$, eq.(\ref{nuda}) naturally suggest that $R(0)$ is a multiple
of the unit matrix. This is usually the case. More precisely, a
solution of the YBE (\ref{ybe}) is called {\it regular} if
\begin{equation}
R(0) = c ~ 1 ~~{\rm that~is}~~  R^{ab}_{cd}(0) = c~
\delta^a_c\;\delta^b_d
\label{reg}
\end{equation}
where $c$ is a non-zero constant.

Setting $\t = 0$ in eqs.(\ref{ybe})
yields with the help of eq.(\ref{reg})
$$
M^{ef}_{ba}(\t') \; \delta^d_c=\delta^f_a \; M^{de}_{cb}(\t')
{}~{\rm where}~~
 M^{ab}_{cb}(\t)\equiv \sum_{1\leq k,l\leq n}R^{ab}_{kl}(-\t)R^{kl}_{cd}(\t)
$$
We thus see that $ M^{ab}_{cb}(\t) $ must have the index structure $
M^{ab}_{cb}(\t) = \delta^a_c\; \delta^b_d \; \rho(\t) $ ,
where $\rho(\t)$ is a c-number function.
This can be written as
$$
 R(\t)R(-\t)=\rho(\t) ~,~{\rm that~is}~~
\sum_{1\leq c,d \leq n} R^{ab}_{cd}(\t) R^{cd}_{ef}(-\t)=
\delta^a_e\;\delta^b_f~\rho(\t)
$$
It follows that $\rho(\t)$
is an even function. This property is usually called `unitarity'
although this may not be always the appropriate name.

{}From eqs.(\ref{monin})-(\ref{trans}) at zero inhomogeneity
 $ \omega_1 = \omega_2 =....= \omega_N = 0 $ and eq.(\ref{reg}), it
follows that
$$
t(0,\{\om_k=0\}) = c^N ~ \Pi_s
$$
where $ \Pi_s $ is the unit shift operator in the horizontal direction.
The momentum operator is then given by
$$
P \equiv -i\log\left[c^{-N} t(0,\{\om_k=0\} )\right]
$$
Moreover, it can be shown \cite{ml,ij} that the operators
$$
{\cal C}_m \equiv \left.
{{\partial^m}\over{\partial\t^m}}\log t(\t,\{\om_k=0\})\right|_{\t=0}
$$
couple $m+1$ neighbor sites on the horizontal line. Usually ${\cal
C}_1$ is a quantum spin chain hamiltonian. The commutativity of
the transfer matrices (\ref{comut}) implies
$$
\left[ {\cal C}_k , {\cal C}_l \right] = 0 ~~\forall ~ k,l
$$
We thus find an infinite number of commuting magnitudes.

We are considering here vertex models where horizontal and vertical
lines are of the same type. This is not necessary for integrability.
It is possible to choose all vertical lines of a given  kind (say with
$n_V$ states per vertical link) and all horizontal lines of a
different kind (say with $n_H \neq n_V$ states per horizontal
link)\cite{ij}. Moreover, it is possible to built integrable models
mixing vertical (or horizontal) lines of different types
\cite{woy}.

It is also possible to construct integrable {\bf face} models where
the variables $(a,b,c,\ldots)$ are attached to the vertices. That is,
to the dual lattice \cite{bax,ijb,car}.

Up to now, we implicitly consider an euclidean two dimensional lattice.
Let us now consider a diagonal-to-diagonal lattice (see fig.4) which
represents a discretized {\bf Minkowski} spacetime in light-cone
coordinates. That is, the axis correspond to $x\pm t$, ($ x $ and $t$
being the usual space and time variables).

In this approach we start from the discretized Minkowski 2D space--time
formed by a regular diagonal lattice of right--oriented and
left--oriented straight lines (see fig. 4). These represent true world--lines
of  ``bare'' objects (pseudo--particles) which are thus naturally divided in
left-- and right--movers. The right--movers have all the same positive rapidity
$\Th$, while the left--movers have rapidity $-\Th$. One can regard $\Th$ as a
cut--off rapidity, which will be appropriately taken to infinity in the
continuum limit. Furthermore, we shall denote by $V$ the Hilbert space of
states of a pseudo--particle (we restrict here to the case in which $V$ is the
same for both left-- and right--movers and has finite dimension $n$, although
more general situations can be considered).

The dynamics of the model is fixed by the microscopic transition amplitudes
attached to each intersection of a left-- and a right--mover, that is to each
vertex of the lattice. This amplitudes can be collected into linear
operators $R_{ij}$, the local $R-$matrices, acting non--trivially only
on the space $V_i\otimes V_j$ of $i$th and $j$th pseudo--particles.
$R_{ij}$ thus represent the relativistic scatterings of left--movers on
right--movers and depend on the rapidity difference $\Th-(-\Th)=2\Th$,
which is constant throughout the lattice. Moreover,
by space--time translation invariance any other parametric dependence
of $R_{ij}$ must be the
same for all vertices. We see therefore that attached to each vertex
there is a matrix $R(2\Th)^{cd}_{ab}$, where $a,b,c,d$ are labels for
the states of the pseudo--particles on the four links stemming out of
the vertex, and take therefore $n$ distinct values (see fig. 1). This is the
general framework of a  vertex model.
The difference with the standard statistical interpretation is
that the Boltzmann weights are in general complex, since we should
require the unitarity of the matrix $R$. In any case,
the integrability of the model is guaranteed whenever $R(\l)^{cd}_{ab}$
satisfy the Yang--Baxter equations (\ref{ybe}).

For periodic boundary conditions, the one--step
light--cone evolution operators $U_L(\Th)$ and $U_R(\Th)$, which act on the
''bare'' space of states $\H_N=(\otimes V)^{2N}$ , ($N$ is the number of
sites on a row of the lattice, that is the number of diagonal lines),
are built from the local $R-$matrices $R_{ij}$ as \cite{npb}.
\begin{eqnarray}
         U_R(\Th)&=& U(\Th)V \;,  \qquad  U_L(\Th) =U(\Th)V^{-1} \nonumber\\
         U(\Th)&=&R_{12}R_{34}\ldots  R_{2N-1\,2N}    \label{evol}
\end{eqnarray}
where $V$ is the one-step space translation to the right.
$U_R$ ( $U_L$ ) evolves states by one step in right (left) light--cone
direction. $U_R$ and $U_L$ commute and their product $U=U_R \, U_L$ is the unit
time evolution operator. The graphical representation of $U$ is given by the
section of the diagonal lattice with fat lines in fig. 4.
If $a$ stands for the lattice spacing, the lattice hamiltonian $H$ and total
momentum $P$ are naturally defined through
\begin{equation}
             U=e^{-iaH} \;,\qquad  U_R\; U_L^{-1}=e^{iaP}    \label{evolu}
\end{equation}
The  action of other fundamental operators is naturally defined on the same
Hilbert space $\H_N$. These are the $n^2$ Yang-Baxter operators for $2N$ sites,
which are conventionally grouped into the $n\times n$ monodromy matrix
$T(\l)=\{T_{ab}(\l),\;a,b=1,\ldots,n\}$.
One usually regards the indices $a,b$ of $T_{ab}$ as {\it horizontal}
indices fixing the out-- and in--states of a reference pseudo--particle.
Then $T(\l)$ is defined as horizontal coproduct of order $2N$ of the
local vertex operators $L_j(\l)=R_{0j}(\l)P_{0j}$, where $0$ label the
reference space and $P_{ij}$ is the transposition in $V_i\otimes V_j$ .
Explicitly
$$
        T(\l) = L_1(\l)L_2(\l) \ldots L_{2N}(\l)
$$
The inhomogenuous generalization $T(\l,{\vec\om\,})$ then reads
$$
  T(\l,{\vec\om\,}) = L_1(\l+\om_1)L_2(\l+\om_2) \ldots L_{2N}(\l+\om_{2N})
$$
and has the graphical representation of fig. 2.
This expression is identical to eq.(\ref{monin}).
$L_j(\l+\om_j)$ can be regarded as the scattering matrix of the $j$th
pseudo--particle carrying formal rapidity $\om_j$ with the reference
pseudo--particle carrying formal rapidity $-\l$.

In the case of our diagonal lattice
of right-- and left--moving pseudoparticles, there exists a specific,
physically relevant choice of the inhomogeneities, namely
$$
\om_k=(-1)^k \Th \;,\quad k=1,2,\ldots 2N
$$
leading to the definition of the {\it alternating} monodromy matrix
$$
      T(\l,\Th)\equiv T(\l,\{\om_k=(-1)^k\Th\})
$$
In fact, the evolution operators $U_L(\Th)$ and $U_R(\Th)$ can be
expressed in terms of the alternating transfer matrix
$t(\l,\Th)=t(\l,\{\om_k=(-1)^k\Th\}) $ as \cite{jpa}
\begin{equation}
      U_R(\Th)= t(\Th,\Th) \;,\quad U_L(\Th)= t(-\Th,\Th)^{-1} \label{ttou}
\end{equation}

Notice that $T(\l,\Th)$ fails to be conserved on the lattice only
because of boundary effects. Indeed from fig. 5, which graphically represents
the insertion of  $T(\l,\Th)$ in the lattice time evolution, one readily sees
that $U$ and  $T(\l,\Th)$ fail to commute only because of the free ends of the
horizontal line. For all vertices in the bulk, the graphical interpretation
of the YB equations (\ref{ybe}), namely that lines can be freely pulled through
vertices, allows to move  $T(\l,\Th)$ up or down, that is to freely commute it
with the time evolution. The problem lays at the boundary: if periodic boundary
conditions are assumed, then the free horizontal ends of  $T(\l,\Th)$ cannot
be dragged along with the bulk, unless they are tied up, to form the
transfer matrix  $t(\l,\Th)$. After all, for p.b.c., the boundary is
actually equivalent to any point of the bulk and thus  $t(\l,\Th)$ commutes
with $U$, as obvious also from eqs.(\ref{ttou}) and the general fact that
$~ [\,t(\l,\Th),\,t(\mu,\Th)\,] = 0 $.
One might think that the thermodynamic limit
$N\to\infty$, by removing infinitely far away the troublesome free ends of
$T(\l,\Th)$, will allow for its conservation and thus for the existence
of an exact YB symmetry with bare $R-$matrix. The situation however is not so
simple: first of all one must fix the Fock sector of the $N\to\infty$
non--separable Hilbert space in which to take the thermodynamic limit.
Different choices leads to different phases with dramatically different
dynamics. Then the non--local structure of  $T(\l,\Th)$ must be taken into
account. It is evident, for instance, that in the spin--wave Fock sector above
ferromagnetic reference states  $T(\l,\Th)$ can never be conserved.
 Indeed, the
working  itself of the Quantum Inverse Scattering Method, where energy
eigenstates are built applying non--diagonal elements of $T(\l,\Th)$ on
a specific ferromagnetic reference state, of course
depends on $T(\l,\Th)$  {\it not} commuting with the hamiltonian!

{F}rom the field--theoretic point of view, the most interesting phase is the
antiferromagnetic one, in which the ground state plays the r\^ole of densely
filled {\it interacting} Dirac sea (this holds for all known integrable lattice
vertex models \cite{npb,jpa,eli,ij}.
The corresponding Fock sector is formed by particle--like
excitations which become relativistic massive particles within the scaling
limit proper of the light--cone approach \cite{jpa}. This consists in letting
$a\to 0$ and $\Th\to\infty$ in such a way that the physical mass scale
\begin{equation}
                \mu=a^{-1} e^{-\kappa\Th}                  \label{mass}
\end{equation}
stays fixed. Here $\kappa$ is a model--dependent parameter which
for the integrable model
where the $R$-matrix is a {\it rational} function of $\t$
takes the general form \cite{eli}
\begin{equation}
            \kappa={{2\pi\, t}\o{h\, s}}                    \label{kconst}
\end{equation}
where $h$ is the dual Coxeter number of the underlying Lie algebra, $s$
equals 1, 2 or 3 for simply, doubly and triply laced algebras,
respectively, and  $t=1 \;(t=2)$  for non--twisted (twisted) algebras.
For the class of model characterized by a trigonometric $R-$matrix
(with anisotropy parameter $\g$) the expression (\ref{kconst}) for $\kappa$
is to be divided by $\g$ \cite{eli}.

The ground state or (physical vacuum) and the particle--like excitations of
this antiferromagnetic phase are extremely more complicated than those of the
ferromagnetic phase. It is therefore very hard to control, in the limit
$N\to\infty$, the action of the
alternating monodromy matrix $T(\l,\Th)$ on the particle--like BA
eigenstates of the alternating transfer matrix  $t(\l,\Th)$.

\end{section}

\begin{section}{ Bootstrap construction of quantum monodromy operators.}

We briefly review in this section the work of refs.\cite{har} where the
exact (renormalized) matrix elements of a quantum monodromy matrix
$\T_{ab}(u)$ ($u$ is the generally complex spectral parameter) were derived
using a bootstrap--like approach for a class of integrable local QFT's. In such
theories there is no particle production  and the  $S-$matrix factorizes. The
two--body $S-$matrix then satisfies the Yang--Baxter equations. Moreover,
in the models considered in  refs.\cite{har} (the $O(N)$ nonlinear sigma
model, the $SU(N)$ Thirring model and the
$0(2N)~({\bar \psi}\psi)^2$  model), thanks to
scale invariance there exist classically conserved monodromy matrices. In
general, the quantum $\T_{ab}(u)$ can be constructed by fixing
its action on the Fock space
of physical in and out many--particle states. The starting point are the
following three general principles:
\begin{enumerate}
 \item
$\T_{ab}(u)$, $a,b=1,2,\ldots,n$, exist as quantum operators and are conserved.
\item
$\T_{ab}(u)$ fulfill a quantum factorization principle.
\item
$\T_{ab}(u)$ is invariant under P, T and the internal symmetries of the theory.
\end{enumerate}
The quantum factorization principle  referred above under 2.
is nowadays called the "coproduct rule". This means that there exists
the following relation between the action of $\T_{ab}(u)$
on $k-$particles states and its action on one--particle states
\begin{eqnarray}
   \T_{ab}(u)\ket{\t_1\a_1,\t_2\a_2,\ldots,\t_k\a_k}_{in}&=
   \sum_{a_1a_2\ldots a_{k-1}} \! \T_{aa_1}(u)\ket{\t_1\a_1}
   \T_{a_1a_2}(u)\ket{\t_2\a_2} \dots \T_{a_{k-1}b}(u)\ket{\t_k\a_k}
\nonumber \\
\\
   \T_{ab}(u)\ket{\t_1\a_1,\t_2\a_2,\ldots,\t_k\a_k}_{out}&=
   \sum_{a_1a_2\ldots a_{k-1}} \!\! \T_{a_1b}(u)\ket{\t_1\a_1}
   \T_{a_2a_1}(u)\ket{\t_2\a_2} \dots \T_{aa_{k-1}}(u)\ket{\t_k\a_k}
\nonumber \label{tin}
\end{eqnarray}
where $\t_j$ and $\a_j \; (1\le j\le k)$ label the rapidities and
the internal quantum numbers of the particles, respectively, in the asymptotic
in and out states. Hence it is understood that $\t_i>\t_j$ for $i>j$.

Although $\T_{ab}(u)$ acts differently on in and out states, the assumption
of conservation is nonetheless consistent. All the eigenvalues of a
maximal commuting subset of $\{\T_{ab}(u),\;a,b=1,2,\ldots,n,\; u\in\CC\}$
are identical for in and out states with given rapidities. Indeed the two in
and out forms of the action on the internal quantum numbers are related by the
unitary permutation
$\ket{\a_1,\a_2,\ldots,\a_k}\to \ket{\a_k,\a_{k-1},\ldots,\a_1}$.

Furthermore, principles 1. and 2. imply that $\T_{ab}(u)$ acts in a trivial
way on the physical vacuum state $\ket{0}$:
$$
           \T_{ab}(u)\ket{0}=\delta_{ab}\ket{0}
$$
This also fixes the normalization of $\T_{ab}(u)$ in agreement with the
classical limit \cite{har}.

An immediate consequence of point 2. is that when $\T_{ab}(u)$ is expanded
in powers of the spectral parameter $u$, it generates an infinite set of
noncommuting and  nonlocal conserved charges. This is the clue to the matching
of the quantum monodromy matrix with its classical counterpart which is written
nonlocally in terms of the local fields.

The main result in refs.\cite{har} was to derive from 1. , 2. and 3. the
explicit matrix elements of $\T_{ab}(u)$ on one--particle states.
This result can be written as
\begin{equation}
           \bra{\t\a}\T_{ab}(u)\ket{\t^\prime \b}= \delta(\t-\t^\prime)
           S^{a\a}_{b\b}(\kappa(u)+\t)                 \label{matel}
\end{equation}
where $S^{a\a}_{b\b}(\t-\t^\prime)$ stands for the $S-$matrix of two--body
scattering
$$
       \ket{\t b,\t^\prime \b}_{in}=\sum_{a\a}\ket{\t a,\t^\prime \a}_{out}
               S^{a\a}_{b\b}(\t-\t^\prime)
$$
and $\kappa(u)$ is an odd function of $u$. Notice that this requires
the presence in the model of particles with indices $a, b, ...$
as internal state labels. In the simplest situation these new labels
coincide with those of the original particles. The appearance of a nontrivial
``renormalization" $u \to \kappa(u)$ is to be expected when there exist
a definition of the spectral parameter outside the bootstrap itself. This is
the case of the models of refs.\cite{har}, which posses Lax pairs and
auxiliary problems which fix the definition of $u$. Here we
adopt the purely bootstrap viewpoint and fix the definition of $u$ so
that $\kappa(u)=u$.
In principle, an extra $u-$ and $\t-$dependent phase factor may
appear in the r.h.s. of eq.(\ref{matel}).
However, no phase showed up in the specific models of refs.\cite{har},
when nonperturbative checks were performed using the operator product
expansion. Eq.(\ref{matel}) can be written in a more suggestive way as
\begin{equation}
          \T_{ab}(u)\ket{\t\b}= \sum_\a\ket{\t\a}
            S^{a\a}_{b\b}(u+\t)               \label{opa}
\end{equation}
This equation, when combined with eqs.(\ref{tin}), completely defines
the quantum monodromy operators in the Fock space. From the YB equations
satisfied by the $S-$matrix it then follows that $\T_{ab}(u)$ fulfills the
YB algebra
\begin{equation}
        {\hat R}(u-v)\left[\T(u)\otimes \T(v)\right]=
\left[\T(u)\otimes \T(v)\right]{\hat R}(u-v)           \label{yba}
\end{equation}
where ${\hat R}^{a\a}_{b\b}(u)=S^{\a a}_{b\b}(u)$.
It should be stressed that the conservation of $\T_{ab}(u)$ implies that
this YB algebra is a true {\it non--abelian infinite symmetry algebra} of the
relativistic local QFT. On the contrary the r\^ole of the YB
algebra in integrable vertex and face models on finite lattices or in
nonrelativistic quantum models is that of a dynamical symmetry
underlying the Quantum Inverse Scattering Method. In these latter cases,
only the transfer matrix, namely
$$
          \tau(u)=\sum_a \T_{aa}(u)
$$
is conserved. Since $\left[\tau(u),\tau(v)\right]=0$, the transfer matrix
just generates an abelian symmetry.

The dynamical symmetry underlying the integrable QFT includes in addition
non--conserved operators $Z_\a(\t)$ which create the particle eigenstates out
of the vacuum. In the bootstrap framework they can be introduced \`a la
Zamolodchikov--Faddeev, by setting
$$
   \ket{\t_1\a_1,\t_2\a_2,\ldots,\t_k\a_k}_{in}  =
   Z_{\a_k}(\t_k)  Z_{\a_{k-1}}(\t_{k-1}) \ldots  Z_{\a_1}(\t_1) \ket0
$$
$$
  \ket{\t_1\a_1,\t_2\a_2,\ldots,\t_k\a_k}_{out}  =
   Z_{\a_1}(\t_1)  Z_{\a_2}(\t_2) \ldots  Z_{\a_k}(\t_k) \ket0
$$
with  the fundamental commutation rules
\begin{equation}
     Z_{\a_2}(\t_2)  Z_{\a_1}(\t_1) =\sum_{\b_1\b_2}
     S_{\a_1\a_2}^{\b_1\b_2}(\t_1-\t_2)  Z_{\b_1}(\t_1) Z_{\b_2}(\t_2)
\label{zzf}
\end{equation}
Combining now eqs.(\ref{tin}),(\ref{opa}) and (\ref{zzf}),
we obtain the algebraic
relation between monodromy and Zamolodchikov--Faddeev operators:
$$
   \T_{ab}(u) Z_\b(\t) = \sum_{c\a}Z_\a(\t)\T_{ac}(u)
                                         S_{b\b}^{c\a}(u+\t)
$$
Together with eqs.(\ref{yba}) and (\ref{zzf}),
these relations close the complete dynamical algebra of an integrable QFT.
For the XXZ spin chain in the regime  $|q| < 1$, the ZF operators have been
identified in ref.\cite{japs} with special vertex operators (or representation
intertwiners of the relevant $q-$deformed affine Lie algebra). They are
uniquely characterized by being solutions of the $q-$deformed
Knizhnik--Zamolodchikov equation and by their normalization \cite{frere}.
\end{section}
\begin{section}{\bf Lattice Construction of Quantum Monodromy
Operators and Bethe Ansatz}

In order to study the infinite volume limit of $T(\l,\Th)$ on the
physical Fock space (that is, finite energy excitations around the
antiferromagnetic vacuum), one needs to compute scalar products of
Bethe Ansatz states to derive relations like (\ref{matel}) or (\ref{opa}) with
 $T(\l,\Th)$ instead of $\T(\l,\Th)$ in the l.h.s. Since this kind of
calculations are indeed possible but rather involved, we computed in
ref.\cite{ybs}   the eigenvalues of  $t(\l,\Th)$ on a generic state of the
physical Fock space. Then, we  compare these eigenvalues with
those of the bootstrap transfer matrix.
This tell us whether the bare and the
renormalized YB algebras have a common abelian subalgebra.
Notice that this fact alone provides a microscopic basis
for the TBA, which originally relies solely on the bootstrap.

We consider once more the sG model as example, although the same
result would apply to any integrable QFT admitting a light--cone
lattice regularization. This class of models contains also the
O(N) nonlinear sigma model and the $SU(N)$ Thirring model considered
from the bootstrap viewpoint in refs.\cite{har}.

The integrable light--cone lattice regularization of
the sG--mT model is provided by the six-vertex model \cite{npb}.
Therefore, the space $V$ is $\CC^2$ and the unitarized
local $R-$matrices can be written
\begin{eqnarray}
       R_{jk}(\l) =&{{1+c}\o2}+{{1-c}\o2}\s^z_j\s^z_k
                     +{b\o2}(\s^x_j\s^x_k + \s^y_j\s^y_k)     \nonumber \\
             b(\l) =&{{\sinh\l}\o{\sinh(i\g-\l)}}\quad ,\quad
             c(\l) ={{\sinh i\g}\o{\sinh(i\g-\l)}}  \quad     \label{rma}
\end{eqnarray}
where $\g$ is commonly known as anisotropy parameter and
$\s^x_j,\s^y_j $ and $ \s^z_j $ are Pauli matrices acting at the site $j$.

The standard Algebrized BA can be applied to the diagonalization of
the alternating transfer matrix $t(\l,\Th)$ with the following
results \cite{npb,jpa,jaca}.
The BA states are written
$$
         \Psi ( {\vec\l}\,) = B(\l_1) .... B(\l_M ) \Omega
$$
where ${\vec\l}\equiv (\l_1,\l_2,\ldots,\l_M)$,
$B(\l_i) = T_{+-}(\l_i + i\g/2 ,\Th)$  and $\Omega$ is the
ferromagnetic ground-state (all spins up).
They are eigenvectors of  $t(\l,\Th)$
$$
t(\l,\Th) \Psi ( {\vec\l}\, ) =
 \Lambda (\l ; {\vec\l}\, )
\Psi ( {\vec\l}\, )
$$
provided the $ \l_i $ are all distinct roots of the ``bare'' BA equations
\begin{equation}
  \left({{\sinh[i\g /2+ \l_j - \Th]}\o{\sinh[i \g /2-\l_j + \Th]} }
     \; {{\sinh[i\g /2+ \l_j + \Th]}\o{\sinh[i \g /2-\l_j - \Th]} }\right)^N =
     - \prod_{k=1}^M
     {{\sinh[+i\g +\l_j - \l_k]}\o{\sinh[-i \g +\l_j -\l_k] }}
                             \label{bae}
\end{equation}
The eigenvalues $ \Lambda (\l ; {\vec\l}\, )$ are the sum of a
contribution coming from
$A(\l)=T_{++}(\l,\Th)$ and one coming from $D(\l)=T_{--}(\l,\Th)$,
\begin{equation}
\Lambda (\l ; {\vec\l}\, ) = \Lambda_A(\l ; {\vec\l}\, )
+ \Lambda_D(\l ; {\vec\l}\, )                    \label{eigev}
\end{equation}
Here
\begin{eqnarray}
    \Lambda_A(\l ; {\vec\l}\, ) &=&
      \exp\left[ -i G(\l , {\vec\l}\, )\right] \nonumber\\
    \Lambda_D(\l ; {\vec\l}\, ) &=&
        e^{-iN \left[ \phi(\l-i\g/2 - \Th , \g /2) +
          \phi(\l-i\g/2 + \Th, \g /2) \right]}
          \exp{[iG(\l - i\g ,  {\vec\l}\, )]}     \nonumber
\end{eqnarray}
and
$$
G(\l,{\vec\l}\,) \equiv \sum_{j=1}^M \phi(\l - \l_j, \g/2) \;,\quad
 \phi(\l,\g) \equiv i\log {  {\sinh(i\g+\l)}\o{\sinh(i\g-\l)}}
$$
$G(\l,{\vec\l}\,)$ is manifestly a periodic function of $\l$
with period $i\pi$.
Notice also that $\Lambda_D (\pm \Th , {\vec \l}) = 0 $. That is, only
$\Lambda_A (\pm \Th , {\vec \l}) $ contributes to
the energy and momentum eigenvalues:
\begin{eqnarray}
         E(\Th ) &=&  a^{-1}\sum_{j=1}^M  \left[
      \phi(\Th+\l_j,\g/2) + \phi(\Th-\l_j,\g/2) - 2\pi \right] \nonumber\\
         P(\Th ) &=&  a^{-1}\sum_{j=1}^M  \left[
      \phi(\Th+\l_j,\g/2) - \phi(\Th-\l_j,\g/2) \right]     \label{enmom}
\end{eqnarray}
The ground state and the particle--like excitations of the light--cone
six--vertex model are well known \cite{npb,ij}: the ground state corresponds to
the unique solution of the BAE with $N/2$ consecutive real roots
(notice that the energy in eq.(\ref{enmom}) is negative definite, so that
the ground state is obtained by filling the interacting Dirac sea). In
the limit $N\to\infty$ this yields the antiferromagnetic vacuum. Holes
in the sea appear as physical particles. A hole located at $\varphi$
carries energy and momentum, relative to the vacuum,
\begin{equation}
   e(\varphi)=2a^{-1}\arctan\left({{\cosh{\pi\varphi/\g}}
          \o{\sinh{\pi\Th/\g}}}\right)          \;,\quad
   p(\varphi)=-2a^{-1}\arctan\left({{\sinh{\pi\varphi/\g}}\o
            {\cosh{\pi\Th/\g}}}\right)   \label{enmomi}
\end{equation}
In the scaling limit $a\to0,\;\Th\to\infty$ with $e(0)$ held fixed, we
then obtain $(e,p)=m(\cosh\t,\sinh\t)$ with
\begin{equation}
 m \equiv 4a^{-1}\exp(-\pi\Th/\g)\;,\quad  \t\equiv-\pi\varphi/\g
\label{massa}
\end{equation}
We have thus proved that the continuum limit is {\bf relativistic} with a
finite non-zero mass. It provides a continuum relativistic massive
field theory out of {\bf any} gapless integrable model. Here, we have
only considered the six-vertex model that yields the massive Thirring
(mT) model \cite{npb}.

We identify $m$
 as the physical mass and $\t$ as the physical rapidity of a sG soliton
(mT fermion) or antisoliton (antifermion). Complex roots of the BAE
are also possible. They correspond to {\it magnons}, that is to
different polarization states of several sG solitons
(mT fermions), or to breather states
(in the attractive regime $\g>\pi/2$).

Within the light-cone approach one can also perform the continuum
limit at the bare level. For the six-vertex model we defined lattice
fermion fields $\psi_n$ and we found their equations of motion on the lattice
Minkowski spacetime \cite{npb}:
\begin{eqnarray}
U_R\psi_{2n-2}U^+_R = U_L\psi_{2n}U^+_L = &{\bar b} \psi_{2n} +  {\bar c}
\psi_{2n-1} + (c -  {\bar c})\psi_{2n}^+\psi_{2n}\psi_{2n-1}- (b +
{\bar b})\psi_{2n-1}^+\psi_{2n-1}\psi_{2n} \nonumber \\
U_L\psi_{2n-1}U^+_L = U_R\psi_{2n+1}U^+_R = & {\bar b} \psi_{2n-1} +  {\bar c}
\psi_{2n} + (c -  {\bar c})\psi_{2n-1}^+\psi_{2n-1}\psi_{2n}- (b +
{\bar b})\psi_{2n}^+\psi_{2n}\psi_{2n-1} \nonumber \\
& {\rm where~} b \equiv b(2\Th) {\rm~and~} c  \equiv c(2\Th) \qquad
\label{eqdes}
\end{eqnarray}
These second quantized field equations are perfectly defined on the
lattice. The bare scaling limit is not identical to the renormalized
limit defined by eq.(\ref{massa}).
The detailed proof in ref.\cite{npb} shows that one finds the bare
continuum mTm if one takes in eq.(\ref{eqdes})
$$
\Th \to \infty \quad , \quad a \to 0 \quad {\rm with}\quad m_0 \equiv {4\o
a}\sin\g~
e^{-2\Th} \quad {\rm kept~ fixed} \label{desli}
$$
We see that the {\bf bare} mass $m_0$ scales as $e^{-2\Th}$ while the
 {\bf renormalized} mass scales as $e^{-\pi\Th/\g}$
[eq.(\ref{massa})].

After some calculations \cite{npb}, the continuum limit of the
momentum and hamiltonian  defined by eq.(\ref{evolu}) take the form
\begin{eqnarray}
P &=& -i \int dx \psi^+\partial_x\psi~~~{\rm and}\nonumber\\
H &=& \int dx \left[ -i
\psi^+\left(\g^5\partial_x+im_0\g^0\right)\psi+{g\o2}\left({\bar
\psi}\g_{\mu}\psi\right)^2 \right] ~~,
\end{eqnarray}
where
\begin{eqnarray}
\psi(x) &=  \left(\begin{array}{r}
\psi_R(x) \\
\psi_L(x) \end{array}\right)~,~\psi_{2n}=\sqrt a\;\psi_R(x+\xi a)~~,~~
\psi_{2n-1}=\sqrt a\;\psi_L(x-\xi a)~\qquad \nonumber\\
{\rm with~}&0<\xi<1/2~,~x=na~,~~{\rm and}~ g
= -2\cot \g ~,~\g_1=-i\s_y~,~\g_0=\s_x~,~\g_5=\s_z~.
\end{eqnarray}
Notice that there is an {\it exact} and finite relation between the bare
continuum ($g$), the lattice  ($\g$) and the renormalized  (${\tilde
\g}={{\g}\o{1-\g/\pi}}$) coupling.

For $R$-matrices acting on finite dimensional spaces $V$ one gets
fermion or parafermion field theories. In order to describe bosonic
field theories one needs infinite dimensional representation spaces
$V$ in the framework of the light-cone approach. [Otherwise, bosons
can appear as bound states of fermions as in the mTm-sG model].

Let us discuss briefly here the rational limit of the six-vertex
$R$-matrix in its spin S representation \cite{krs}.
That is, for $V = \CC^{2S+1}$.
\begin{equation}
R_{jk}(\t) =
{{\Gamma(2S+1+i\t)\Gamma(J+1-i\t)}\o{\Gamma(2S+1-i\t)\Gamma(J+1+i\t)}}
\label{spins}
\end{equation}
where the operator $J$ is defined by
$$
J(J+1) = 2S(S+1) + 2 {\vec S}_j \otimes  {\vec S}_k
$$
where ${\vec S}_j$ and ${\vec S}_k$ are spin $S$ operators acting on
the spaces $V_j$ and $V_k$ respectively. [ $({\vec S}_j)^2 = ({\vec
S}_k)^2= S(S+1) $]. The hamiltonian and momentum (\ref{evolu})
describe in the $S = \infty$ limit the principal chiral model (PCM)
\cite{fr,ddv}. However, this is not the full hamiltonian. One finds in
this way states which are left (or right) $SU(2)$-singlets. The lattice
current construction (\ref{gener})-(\ref{corcl}) [see below] holds
for the  PCM. Notice that for large $\t$ the $R$-matrix (\ref{spins})
possess an expansion like eq.(\ref{rasi}). Then, the whole procedure
works yielding a conserved a curvatureless current that we can
identify  either with the  $SU(2)_L$ or with the  $SU(2)_R$ current.
This whole construction generalizes to the $SU(N)$ PCM. It also
generalizes to PCM with one anisotropy axis (trigonometric
$R$-matrices)\cite{kir}.

The light-cone approach to the sine-Gordon-mTm  model using bosonic fields
is worked out in ref.\cite{favo}.

\end{section}
\begin{section}{\bf Vertex Models and Field Theories associated to
q-deformed Lie Algebras}

The $R$-matrices solutions of the YB eq.(\ref{ybe}) can be classified
according to
\begin{enumerate}
 \item The Lie algebra (or q-deformed  Lie algebra) to whom they are
associated.
\item  The couple of Lie algebra representations where they act:
 $V \otimes V'$, ($~V$may coincide or not with $V'$) .
\end{enumerate}
The six-vertex $R$-matrix is associated to the $q-A_1$ Lie algebra in
its fundamental (spin 1/2) representation. The $R$-matrix \cite{ij}
\begin{eqnarray}
R^{ab}_{ab}(\theta)&=&\frac{\sin\gamma}{\sin(\gamma-\theta)}~
e^{i\theta \,{\rm sign}(a-b)}\;\; ,\;\; a\neq b\nonumber \quad ;\\
R^{ab}_{ba}(\theta)&=&\frac{\sin\theta}{\sin(\gamma-\theta)}\;\;,\;\;a\neq
b \quad ; \label{mrtri}\\
R^{aa}_{aa}(\theta)&=&1\nonumber\\
&&1\leq a,b\leq n\nonumber
\end{eqnarray}
corresponds to the $q-A_{n-1}$ Lie algebra in its $n$-dimensional (quark)
representation. When $ q = e^{i\g} $ is a root of unity
representations which are unknown for $q=1$ appear and hence new models
can be constructed. They are better defined in face language. For the $q-A_1$
case they are called RSOS models\cite{bax}. $R$-matrices associated to
other q-Lie algebras can be found in \cite{sem}.

The eigenvectors of the transfer matrix can be obtained {\it via} the
Bethe Ansatz (BA) in its various generalizations. (The BA can also be
formulated in face language \cite{car,ijb}). When the $R$-matrix
corresponds to a Lie algebra of rank larger than one, the
nested Bethe Ansatz (NBA)must be used.

The nested Bethe Ansatz (NBA) is probably the most sophisticated
algebraic construction of eigenvectors for integrable lattice models.
It consists of several levels of BA each one inside the previous one.
This is the reason of its name. The NBA has been worked out for the
 $A_{n-1}$ trigonometric and hyperbolic vertex model
\cite{ij},  for the $Sp(2n)$  symmetric vertex model
 \cite{kr} and for $O(2n)$ symmetric vertex model \cite{mk} (always in
the fundamental representation).
The structure of the NBA is closely related to the respective Dynkin
diagram. For $D_n$ , one starts by one end of the `fork', goes till
the end of the diagram and then back till the other end of
the `fork' \cite{mk}.

The NBA equations (NBAE) for a class of vertex models associated
to simple Lie Algebras
has been proposed in ref.\cite{nik} and solved (in a large extent)
in ref.\cite{eli}.  Let us summarize the more relevant results for the
field theory limit.

The structure of the NBAE for a given $R$-matrix is dictated by the
associated Dynkin diagram. There are sets of NBA roots associated to each
spot of the  Dynkin diagram. The NBAE couple the roots associated to
the same spot and to the roots associated to spots connected to it
in the Dynkin diagram. The structure of the ground state may be
antiferromagnetic (AF) or ferromagnetic (F) depending on the chosen
regime (values of $\t$ and $\g$ ).

The AF ground state yields the more interesting field theories. It is
formed by filling all `Dirac' seas with BA roots. There is a Dirac sea for
each NBA level.  The ground state roots are  real for simply laced Lie
algebras apart of a constant imaginary part that depends on the level in a
simple way \cite{ij,eli}. For non-simply laced cases \cite{eli} and for
non-fundamental representations of all algebras, complex (`string' type)
roots form the ground state.

On the top of the AF ground state (renormalizes or physical vacuum)
there are excitations. One finds as many branches of excitations as
the rank of the Lie algebra. They follow making holes and adding
complex roots to each Dirac sea. The mass spectrum is usually q-independent
except for non-compact q-Lie algebras \cite{ps}. The mass spectrum coincides
(for simply laced cases) up to a general factor with the components of the
Perron-Frobenius eigenvector of the Cartan  matrix.

The field theories obtained for $R$-matrices in the fundamental
representations are basically fermions or para-fermion models.
In ref.\cite{eli} we computed the mass spectrum and the S-matrix in
these models for most of the q-deformed simple Lie algebras.

Let us now briefly discuss the scaling limit of vertex models with
rational $R$-matrices associated to a Lie algebra ${\cal G}$.
These  $R$-matrices have the asymptotic behaviour
\begin{equation}
R(\t) \buildrel{\t\to\infty}\over= P\left[ 1 + {{\Pi + \mu}\o{i\t}}+
O({1\o{\t^2}})\right] \label{rasi}
\end{equation}
where $\mu$ is a numerical constant, $P^{ab}_{cd} =
\delta^a_d\;\delta^b_c $ is the exchange operator and
\begin{equation}
\Pi=\sum_{\a=1}^{dim{\cal G}}T_{\a}\otimes T^{\a}
\label{casi}
\end{equation}
We then introduce the lattice operator
\begin{equation}
T^{\a}_n\equiv 1 \otimes \ldots \otimes
\overbrace{T^{\a}}^{n^{th.} site}\otimes \ldots \otimes 1
\label{gener}
\end{equation}
Using eqs.(\ref{evol}),(\ref{rasi}) and (\ref{casi}) and the Lie algebra
commutators
$$
\left[T^{\a},T^{\b}\right] = i f^{\a\b}_\g ~ T^{\g}
$$
we can show that the operators $ T^{\a}_n $ obey  {\it local}
equations of motion on the lattice \cite{jpa}
\begin{eqnarray}
U_R T^{\a}_{2n-2} U^+_R = U_L T^{\a}_{2n} U^+_L =& T^{\a}_{2n} +
{{2i}\o{\t}} f^{\a\b}_\g  T^{\b}_{2n-1} T^{\g}_{2n}+O({1\o{\t^2}}) ,
\nonumber\\
U_R T^{\a}_{2n-1} U^+_R = U_L T^{\a}_{2n+1} U^+_L =& T^{\a}_{2n-1} -
{{2i}\o{\t}} f^{\a\b}_\g  T^{\b}_{2n-1} T^{\g}_{2n}+O({1\o{\t^2}}) .
\label{coret}
\end{eqnarray}
The bare scale limit is now defined as $a \to 0, ~ \t \to \infty, ~ x
= na$ fixed. We find
\begin{equation}
\partial^{\mu}J_{\mu}^{\a} (x) = 0 \quad, \quad \partial_0 J^{\a}_1
- \partial_1 J^{\a}_0 + ig  f^{\a\b}_\g \left[J^{\b}_0,J^{\g}_1\right]
= 0 . \label{eqco}
\end{equation}
where $\mu = 0, 1$ and in light-cone coordinates
\begin{equation}
 J^{\a}_R \equiv {1 \o {ga\t}}~ T^{\a}_{2n}\quad , \quad
 J^{\a}_L \equiv {1 \o {ga\t}} ~ T^{\a}_{2n-1}
\label{corcl}
\end{equation}
Therefore we have a lattice version of the  ${\cal G}$-algebra currents
$J^{\a}_{\mu}(x)$ associated to an exactly integrable discretization
of the field theory model. Eqs.(\ref{eqco}) characterize the currents
in the non-abelian Thirring model associated to the Lie algebra
${\cal G}$. This model has as Lagrangian
\begin{equation}
{\cal L} = i {\bar \psi} \not \! \partial \psi - {g\o 4}\left( {\bar
\psi}\g_{\mu}
 T^{\a} \psi\right)\left( {\bar \psi}\g^{\mu} T^{\b} \psi\right)
 K_{\a\b} \label{lagr}
\end{equation}
Here $\psi$ transforms under an irreducible representation $\rho$ of
 ${\cal G}$, $ T^{\a} $ are the  ${\cal G}$-generators in that
representation and $ K_{\a\b} $ is proportional to the inverse of the
Killing form. Actually the hamiltonian and momentum [H and P defined
by eq.(\ref{evolu})] describe the zero-chirality (massive) sector of the model
(\ref{lagr}) (see ref.\cite{jpa}) and we can identify
\begin{equation}
J^{\a}_{\mu}(x) = {\bar \psi}\g_{\mu} T^{\a} \psi
\end{equation}
The renormalized scaling limit is discussed in \cite{jpa,ij} and
through eqs.(\ref{mass})-(\ref{kconst}).
\end{section}
\begin{section}{Thermodynamic limit of the transfer matrix from the
Bethe Ansatz}

In ref.\cite{ybs}  it is  shown that the bootstrap construction
(discussed in sec.2) of conserved
$\T_{ab}(u)$ generalizes to integrable models with trigonometric
$R-$matrices such as the sine-Gordon or massive Thirring model.
In such  cases  the classical limit is abelian, as shown explicitly
in ref.\cite{ybs}.

The main aim of ref.\cite{ybs} was  to investigate and clarify, from a
microscopic point of view, the problem of unveiling the existence
of the infinite YB symmetry of the sG--mT model. In other words, since
lattice models provide regularized version of QFT, we seek an explicit
connection between the lattice and the bootstrap YB algebras.

In order to investigate the operators present in such QFT,
it is important to learn how the monodromy operators $T_{ab}(\l, \Th)$ act
on physical states. In ref.\cite{ybs} we explicitly compute
the eigenvalues of the alternating six--vertex transfer matrix
$t(\l,\Th) $, on a generic $n-$particle state,
in the thermodynamic limit.

The eigenvalues of $t(\l,\Th)$
turned out to be $i\pi-$periodic and multi--valued functions of $\l$,
each determination of  $t(\l,\Th)$ being a meromorphic function of
$\l$. We call $t^{II}(\l,\Th)$ and $t^{I}(\l,\Th)$ the determinations
associated with the periodicity strips closer to the real axis .
 The ground--state
contribution $\exp[-iG(\l)_V]$ is exponential on the lattice size, as
expected, whereas the excited states contributions are finite and
express always in terms of hyperbolic functions

In strip I $|\Im \l| < \g/2 $, we define the
renormalized type I transfer matrix
\begin{equation}
    t^I(\l)=  \lim_{N\to\infty} t(\l,\Th)
             \exp\bigl[iG^I(\l)_V\bigr](-)^{J_z-N/2}  \label{rentm}
\end{equation}
where $J_z=N/2-M$ is to be identified with the soliton (or fermion) charge
of the continuum sG--mT model. The last sign factor in eq.(\ref{rentm})
 corresponds
to square--root branch choice suitable to obtain the relation
\begin{equation}
      t^I(\pm\Th)= \exp\{-ia[P_{\pm}-(P_{\pm})_V]\}   \label{tppm}
\end{equation}
where $ P_{\pm} \equiv (H\pm P)/2$ [see eqs.(\ref{evolu}), (\ref{ttou}),
(\ref{enmom})] and
$(P_{\pm})_V$ stands for the vacuum contribution.
Notice that the $\Th-$dependence of  $t^I(\l)$ has been completely
canceled out, since it is present only in the vacuum contribution. In fact,
from the Bethe Ansatz calculations in ref.\cite{ybs}, we read the eigenvalue
$\Lambda^I(\l)$ of $t^I(\l)$ on a generic particle state:
\begin{equation}
     \Lambda^I(\l)= \exp\left[-2i\sum_{n=1}^k \arctan
              \left(e^{\pi\l/\g+\t_n} \right)\right] =  \prod_{n=1}^k
       \coth\left({{\pi\l}\o{2\g}}+{{\t_n}\o2}+{{i\pi}\o 4}\right)
\label{eig}
\end{equation}
where $\t_n\equiv -\pi\varphi_n/\g$ are the physical particle rapidities.
Suppose now we expand  $\log{\Lambda^I}(\l)$ in
powers of $z=e^{-\pi|\l|/\g}$ around $\l=\pm\infty$,
$$
  \pm i \log\Lambda^I(\l) =   \sum_{j=0}^\infty z^{2j+1}
                {{(-1)^j}\o{j+1/2}} \sum_{n=1}^k e^{\pm(2j+1)\t_n}
$$
One has to regard the coefficients of the expansion parameter $z$ as the
eigenvalues of the conserved abelian charges generated by the transfer matrix.
The additivity of the eigenvalues implies the locality of the charges.
In terms of operators we can write, around $\l=\pm\infty$,
\begin{equation}
    \pm i \log t^I(\l)  = \sum_{j=0}^\infty
                 \left[{{4z}\o m}\right]^{2j+1} I_j^\pm  \label{expaop}
\end{equation}
where $I_0^\pm = p_{\pm} $ is the continuum light--cone
energy--momentum and
the  $I_j^\pm$, $j \ge 1$, are local conserved charges  with dimension $2j+1$
and  Lorentz spin $\pm(2j+1)$. Their eigenvalues
$$
       {{(-1)^j}\o{j+1/2}} \sum_{n=1}^k\left[{m\o4}e^{\pm\t_n}\right]^{2j+1}
$$
coincide with the values on multisoliton solutions of the
higher integrals of motion of the sG equation \cite{leon}.
It is remarkable that these eigenvalues are free
of quantum corrections although the corresponding operators in terms of
local fields certainly need renormalization. Let us stress that
explicit expressions for these conserved charges can be obtained
by writing the local $R-$matrices in terms of fermi operators, as in
ref.\cite{npb}. Notice also that, combining
eqs.(\ref{tppm}) with (\ref{expaop}), and recalling the scaling law
(\ref{massa}), we can write
\begin{equation}
     P_{\pm}-(P_{\pm})_V = p_{\pm}+{m\o4}\sum_{j=1}^{\infty}
               \left({{ma}\o4}\right)^{2j} I_j^\pm      \label{hmpl}
\end{equation}
That is, the light-cone lattice hamiltonian and momentum can be expressed in a
precise way as the continuum hamiltonian and momentum plus an infinite series
of continuum higher conserved charges, playing the r\^ole of irrelevant
operators.

The explicit  Bethe Ansatz calculation in ref.\cite{ybs}
showed that in the strip II, the lattice transfer matrix eigenvalues
match  with the bootstrap transfer matrix eigenvalues.

We obtained as  general form of the $A$ and $D$ contributions to the
eigenvalue of $t(\l,\Th)$ [see eq.(\ref{eigev})]
on the $N\to\infty$ limit of the BA states for
$\l$ in strip II\cite{ybs}:
$$
     \Lambda_A(\l) = -e^{-iG^{II}(\l)_V} \left\{ \prod_{n=1}^k
         S(x_n) \coth{{x_n}\o2}  \right\}
        \prod_{j=1}^m
             {{\sinh {\hat\g} [i/2+({\pi\o\g}(\l+i\g/2)-u_j)/\pi]}  \o
              {\sinh {\hat\g} [i/2-({\pi\o\g}(\l+i\g/2)+u_j)/\pi]}}
$$

and $$
 \Lambda_D(\l) = -e^{-iG^{II}(\l)_V} \left\{ \prod_{n=1}^k
        S(x_n){\hat b}(x_n) \coth{{x_n}\o2}\right\}
        \prod_{j=1}^m
             {{\sinh {\hat\g} [3i/2+({\pi\o\g}(\l+i\g/2)-u_j)/\pi]} \o
              {\sinh {\hat\g} [-i/2+({\pi\o\g}(\l+i\g/2)-u_j)/\pi]}}
$$
where for definiteness we chose the strip II , $-\pi+\g/2<\Im\l<-\g/2$ and set
$x_n={\pi\o\g}(\l+i\g/2)+\t_n$.
 The distinct numbers $u_1,u_2,\ldots,u_m$ must satisfy
the BA equations
$$
     \prod_{n=1}^k
           {{\sinh {\hat\g} [i/2+(u_j+\t_n)/\pi]} \o
            {\sinh {\hat\g} [i/2-(u_j+\t_n)/\pi]}} =
     -\prod_{r=1}^m
           {{\sinh {\hat\g} [+i+(u_j-u_r)/\pi]}  \o
            {\sinh {\hat\g} [-i+(u_j-u_r)/\pi]}}
$$
These last two expressions can be connected  with that for the eigenvalues of
the bootstrap transfer matrix $\tau(u)$ \cite{ybs}, provided we {\it identify}
$u$ with ${\pi\o\g}(\l+i\g/2)$. We find indeed \cite{ybs}:
\begin{equation}
          \Lambda(\l) = -e^{-iG^{II}(\l)_V} \xi(u) \prod_{n=1}^k
          \coth\left({{u+\t_n}\o2}\right)   \label{landxi}
\end{equation}
where $\xi(u)$ is the eigenvalue of the  bootstrap transfer matrix
$\tau(u)$ and  $\l$ is in strip II .
In analogy with eq.(\ref{rentm}), we now define the type
II renormalized transfer matrix
$$
  t^{II}(\l) =  \lim_{N\to\infty} t(\l,\Th)
        \exp\bigl[iG^{II}(\l)_V\bigr](-)^{J_z-N/2}
$$
Then, taking into account eq.(\ref{eig}), eq.(\ref{landxi}) can be rewritten
\begin{equation}
  \xi(u) =  {{\Lambda^{II}\left({\g\o\pi}u - i{\g\o2}\right)}
              \o {\Lambda^I\left({\g\o\pi}u - i{\g\o2}\right)}}
\label{xilan}
\end{equation}
Notice that the dependence on the cutoff rapidity $\Th$
has completely disappeared from the r.h.s. of eq.(\ref{xilan}).
This holds true both
for the explicit dependence in the vacuum function $G(\l)_V$ and
for the implicit dependence through the bare BAE, which are now
replaced by the $\Th-$independent higher--level ones. In other words, the
eigenvalues of the bootstrap transfer matrix can be recovered from the
light--cone regularization already on the infinite diagonal lattice,
with no need to take the continuum limit.
This should cause no surprise, since after all a factorized
scattering can be defined also on the infinite lattice, with physical
rapidities replaced by lattice rapidities [see eq.(\ref{enmomi})].
The bootstrap
construction of the quantum monodromy operators $\T_{ab}(u)$ then proceeds just
like on the continuum. In this case, some $q_0-$deformation of the two
dimensional Lorentz algebra should act as a symmetry on the physical
states. This $q_0$ becomes unit when $\Th \to \infty$.

We then compare these Bethe Ansatz eigenvalues with the eigenvalues of
the bootstrap transfer matrix $\tau(u)$.
Remarkably enough, we find the following simple relation
between the two results, for $0<\g<\pi/2$ (repulsive regime),
\begin{equation}
     \tau(u)= t^{II}\bigl({\g\o\pi}u-i{\g\o2},\Th\bigl)
              \,t^I\bigl({\g\o\pi}u-i{\g\o2},\Th\bigl)^{-1}
\label{gorda}
\end{equation}
where $t^{II}(\l,\Th)$ and $t^{I}(\l,\Th)$ have been normalized to one
on the ground state .
 Thus, we succeed in connecting the bootstrap  transfer matrix
$\tau(u)$ of the sG-mT model with the alternating transfer matrix  $t(\l,\Th)$
of the six vertex model. In the thermodynamic limit $\tau(u)$ coincide
with the jump between the two main determinations of $t(\l,\Th)$ . Notice the
renormalization of the rapidity by $ \g/ \pi$ and the precise overall shift
by $i\g/2$ in the argument in order the equality to hold.

We find in addition that $t(\l,\Th)$, for $ 0 < \Im \l < \g/2$, generates
the hamiltonian and momentum
together with an infinite number of higher--dimension and higher--spin
conserved abelian charges, through expansion in powers of $e^{\pm\pi\l/\g}$.
We see therefore that the same bare operator generates two kinds of conserved
quantities. Energy and momentum as well the higher--spin abelian charges are
local in the basic fields which interpolate physical particles, whereas the
infinite set of charges obtained from the jump from
 $t^{II}(\l,\Th)$ to $t^{I}(\l,\Th)$
  are nonlocal in the same fields. The
fact that local and nonlocal charges
come from different sides of a natural boundary, clearly
shows that they carry independent information. That is, one cannot
produce the nonlocal charges from the sole knowledge of the local
charges. We also recall that the monodromy matrix $T(\l,\Th)$
can be written in terms of the lattice Fermi fields of the mT model
\cite{npb}, so that local and non local charges do admit
explicit expressions in terms of local field operators.

We expect eqs.(\ref{gorda}) and (\ref{hmpl}), and the discussion
below eq.(\ref{gorda}) , to be valid for many other integrable models
provided the appropriate rapidity renormalization and imaginary
shift are introduced.

The quantum monodromy operators $\T_{ab}(u)$ generate a Fock representation
of the $q-$deformed affine Lie algebra $U_q({\hat\G})$
corresponding to the given
$R-$matrix. More precisely, by expanding $\T_{ab}(u)$ in powers of $z=e^u$
around $z=0$ and $z=\infty$, one obtains non--abelian non-local
conserved charges representing the algebra  $U_q({\hat\G})$ on the
Fock space of in-- and out--particles. This connects our approach
based on the YB symmetry, to the
$q-$deformed algebraic approach of ref.\cite{leclair}.
$U_q({\hat\G})$ is a Hopf algebra endowed with an universal
$R-$matrix, which reduces to the $R-$ explicitly entering the YB algebra,
upon projection to the finite--dimensional vector space spanned by the
indexes of $\T_{ab}(u)$ \cite{frere}. In particular, the two expansions around
$z=0$ and $z=\infty$ generate the two Borel subalgebras of  $U_q({\hat\G})$.
A single monodromy matrix  $\T(u)$ is sufficient for this purpose, since this
field--theoretic representation has level zero . This fact receives a
new explanation in the light--cone approach, since  $U_q({\hat\G})$
emerges as true symmetry only in the infinite--volume limit above the
antiferromagnetic ground state (with no need to take the continuum limit),
but its action is uniquely defined already on {\it finite} lattices, and
all finite--dimensional representations have level zero.

\end{section}

\vskip 12pt
\begin{center}
{\bf Figure Captions}
\end{center}
\vskip 12pt
\begin{description}
\item{Fig.1.}
The $R$-matrix elements $R^{ab}_{cd}(\t)$ define the
statistical weight of the depicted vertex configuration
\item{Fig.2.}
Graphical representation of the inhomogeneous monodromy matrix. The angles
between the horizontal and the vertical lines are site--dependent in an
arbitrary way.
\item{Fig.3.}
The Yang-Baxter equation.
\item{Fig.4.}
Light--cone lattice representing a discretized portion of
Minkowski space--time. A $R-$matrix of probability amplitudes is attached
to each vertex. The bold lines correspond to the action, at a given time,
of the one--step evolution operator $U$.
\item{Fig.5.}
Insertion of the alternating monodromy matrix in the light--cone
lattice.
\item{Fig.6.}
The two main determinations, $G^I(\l)$ and $G^{II}(\l)$ are defined by
$G(\l)$ with $\l$ in strips I and II, respectively.
 \end{description}
\end{document}